\documentclass[%
reprint,
showpacs,preprintnumbers,
nofootinbib,
 amsmath,amssymb,
 aps,
floatfix,
]{revtex4-1}

\usepackage{graphicx}% Include figure files
\usepackage{dcolumn}% Align table columns on decimal point
\usepackage{bm}% bold math
\usepackage[colorlinks=true,urlcolor=blue, linkcolor=blue, citecolor=blue]{hyperref}% add hypertext capabilities
\usepackage{amsthm}

\usepackage[inline]{enumitem}  % Inline lists
\usepackage{makecell}
\usepackage{microtype}
\usepackage[caption=false]{subfig}
\usepackage[color=orange!40]{todonotes}
\usepackage{xcolor}  % To color our equations

\makeatletter  % Coloring in equations
\def\mathcolor#1#{\@mathcolor{#1}}
\def\@mathcolor#1#2#3{%
  \protect\leavevmode
  \begingroup
    \color#1{#2}#3%
  \endgroup
}
\makeatother

\renewcommand{\say}[1]{``#1''}

\DeclareMathOperator{\E}{{\mathbb{E}}}

\newcommand{\model}[1]{\ensuremath{\mathbb{M}_{\text{#1}}}}
\newcommand{\modelone}[1]{\ensuremath{\mathbb{M}^1_{\text{#1}}}}

\newcommand{\mi}{I}
\newcommand{\nmi}{\mathrm{NMI}}
\newcommand{\rnmi}{\mathrm{rNMI}}
\newcommand{\rrnmi}{\mathrm{rrNMI}}
\newcommand{\cnmi}{\mathrm{cNMI}}
\newcommand{\ami}{\mathrm{AMI}}

\newcommand{\baseline}{B}
\newcommand{\bound}{M}

\newcommand{\true}{\ensuremath{\mathcal{T}}}
\newcommand{\pred}{\ensuremath{\mathcal{C}}}
\newcommand{\size}[1]{\left|#1\right|}

\newcommand{\calC}{\mathcal{C}}

\newcommand{\calT}{\mathcal{T}}

\newtheorem{theorem}{Theorem}

\begin{document}

\preprint{Preprint}

\title{Metrics matter in community detection}

\author{Arya D. McCarthy}
 \email{arya@jhu.edu}
 \altaffiliation[Partially conducted at ]{Southern Methodist University.}%Lines break automatically or can be forced with \\
\author{Tongfei Chen}%
\author{Rachel Rudinger}%
\affiliation{%
 Department of Computer Science \\
 Johns Hopkins University
}%

\author{David W. Matula}
\affiliation{
 Department of Computer Science and Engineering\\
 Southern Methodist University
}%

\date{\today}

\begin{abstract}

We present a critical evaluation of normalized mutual information (NMI) as an evaluation metric for community detection.
NMI exaggerates the leximin method's performance on weak communities:
Does leximin, in finding the trivial singletons clustering, truly outperform eight other community detection methods?
Three NMI improvements from the literature are AMI, rrNMI, and cNMI.
We show equivalences under relevant random models, and \textbf{for evaluating community detection, we advise one-sided AMI under the $\model{all}$ model \textmd{(all partitions of \(n\) nodes)}}.
%We also show that rrNMI is the one-sided version of cNMI in all random models.
This work seeks
\begin{enumerate*}[label=(\arabic*), after=\unskip{.}, itemjoin={{; }}, itemjoin*={{, and }}]
    \item to start a conversation on robust measurements
    \item to advocate evaluations which do not give \say{free lunch}
\end{enumerate*}

\end{abstract}

\maketitle

\section{Introduction}
Unsupervised algorithms---like those for community detection (CD), investigated in many discipline for half a century \citep{matula1977graph}---present a challenge for appraisal. In CD, we circumvent the problems of intrinsic measures by using external evaluation tasks. Practitioners apply CD methods to benchmark graphs containing \say{ground truth} communities, then compute an agreement measure to determine how well those communities are recovered. These measures differ from typical classification accuracy because there are no specific labels (e.g. no notion of ``Cluster 2'')---only groups of similar entities.

The popular measure in CD is normalized mutual information (NMI). Its theoretical flaws have been noted~\cite{ lai2016corrected, peel2017ground, vinh2010information, zhang2015relationship}. Particularly relevant is the non-homogeneity of the measure: NMI awards credit for low-information guessing~\cite{peel2017ground}. This deficiency has demonstrable implications for method selection, which we later show using the leximin method as an example.

%Improvements to NMI have gained little traction in CD\@. 
A sequence of proposed improvements \citep{zhang2015evaluating, zhang2015relationship} in the CD community led to the recent \emph{corrected NMI} (cNMI)~\cite{lai2016corrected}. A common, older measure, \emph{adjusted mutual information} (AMI)~\cite{vinh2010information}, has garnered recent attention in CD~\cite{mccarthy2017gridlock, peel2017ground}. AMI augments NMI's consistent upper bound (1.0) with a consistent zero expectation to adjust for chance clusterings:%
\footnote{Negative AMI indicates worse-than-chance clusterings.}
%\todo{Switch all notation from \(C\) and \(\hat{C}\) to \(\mathcal{T}\) and \(\calC\).}
%\[
%\mathrm{AMI}_{\mathit{\mathcolor{magenta}{sum}}}(\hat{C}, C) = \frac{\mathrm{MI}(\hat{C}, C) \mathcolor{blue}{- \E_{\hat{C}', C'}\left[ \mathrm{MI}(\hat{C}', C') \right] }}{\mathcolor{magenta}{\frac{1}{2}} \left(H(\hat{C}) \mathcolor{magenta}{+} H(C)\right) \mathcolor{blue}{- \E_{\hat{C}', C'}\left[ \mathrm{MI}(\hat{C}', C') \right] }}
%\textrm{.}
%\]
% \begin{equation}
%   \label{eq:intro-ami}
% \!\!\!  \ami_{\mathcolor{magenta}{0}}^{\mathcolor{blue}{\rm perm}}(\pred, \true) = 
% \frac%
% 	{\mi(\pred, \true) \mathcolor{blue}{- \E_{\pred', \true'}\left[ \mi(\pred', \true') \right] }}%
% 	{\mathcolor{magenta}{\sqrt{\mathcolor{black}{H(\pred)} \cdot \mathcolor{black}{H(\true)}}} \mathcolor{blue}{- \E_{\pred', \true'}\left[ I(\pred', \true') \right] }}
% \textrm{.}
% \end{equation}
\begin{equation}
  \label{eq:intro-ami}
\!\!\!  \ami(\pred, \true) = 
\frac%
	{\mi(\pred, \true) \mathcolor{blue}{ - \baseline(\pred, \true)}}%
	{\mathcolor{magenta}{\bound(\pred, \true)} \mathcolor{blue}{ -  \baseline(\pred, \true)}}
\textrm{,}
\end{equation}
\noindent where $\baseline(\pred, \true)$ is a baseline function that is used to adjust the metric to zero expectation, and $\bound(\pred, \true)$ guarantees consistent upper bound of the metric. In previous literature, one common incarnation of these these functions could be
\begin{align}
\baseline(\pred, \true) &= \E_{\pred', \true'}\left[ \mi(\pred', \true') \right] \label{eqn:baseline} \ ; \\ 
\bound(\pred, \true) &= \sqrt{H(\pred) \cdot H(\true)} \ . \label{eqn:bound}
\end{align}
 %Many forms of NMI and derived measures (e.g. AMI) exists. % We color-code the averaging method in magenta. Including the expectation model, in blue, differentiates AMI from NMI.
Many variants exist, by changing the definitions of the $B$ and $M$ functions.

Note the expectation operator in \autoref{eqn:baseline}. Over what distribution is the expected value computed? This is called \emph{random models} in recent work \cite{gates2017impact}.
The literature has implicitly computed expectations over \(\mathbb{M}_\mathrm{perm}\): all partitions of the same \emph{class} (or cluster-size pattern) as the observation \citep{zhang2015evaluating,zhang2015relationship,lai2016corrected,vinh2009information,vinh2010information}.
We argue that more appropriate for CD is an \textbf{expectation over $\model{all}$ instead of $\model{perm}$}: all partitions of \(n\) nodes. We also advise \textbf{one-sided random models} for comparing against a fixed ground truth.

The main contributions of this work are
\begin{itemize}[before=\unskip{: }]
    \item With the \emph{leximin method} as an example, we show the need for an improved evaluation metric in CD\@.%~ (\autoref{sec:experiment})
    \item We identify an evaluation function which better matches the community detection problem domain.
    \item We advocate for the use of the adjusted metric AMI\@, with slight modifications from its proposed form.
    \item We provide thorough analysis of the relationships between AMI and other evaluation functions.
\end{itemize}

% \clearpage

%
%
%
%
%
\section{Community Detection}

A number of tasks on graphs ask that you partition the graph's nodes to maximize a score function.
Situated between the microscopic, node level and the macroscopic, whole-graph level, these partitions form a \emph{mesoscopic structure}---be it a core--periphery separation, a graph coloring, or our focus: community detection.
Community detection has been historically ill-defined, though the intuition is to collect nodes with high interconnectivity (or \emph{edge density}) into communities with low edge density between them.
The task is analogous to clustering, which groups points that are near one another in some fashioned metric space.

To bring some rigor to the task, we often choose \emph{modularity} \cite{newman2006modularity} as the objective function.
%:
%\begin{equation}
%    Q = \frac{1}{4 \left|E\right|} \sum_{i,j} \left(A_{ij} - \frac{\deg i \deg j}{2 \left|E\right|} \right) \delta_{\pi_i \pi_j}
%    \text{,}
%\end{equation}
%where we use the Kronecker delta, the adjacency matrix~\(\boldsymbol{A}\)\@, and \(\boldsymbol{\pi}\) which gives the community label of a node.
Modularity balances the objectives of many edges within clusters (the first term) and avoiding giant, all-encompassing communities (the second term).
Directly maximizing modularity is \textsf{NP}-hard \citep{brandes2006maximizing}, as is approximating it to within any constant factor \citep{dinh2015network}.
For this reason, most community detection methods are heuristics that approximate modularity.

Still, modularity isn't the be-all, end-all in community detection. 
In real-world data, we may have a notion of the communities already, and in these cases we only value modularity insofar as it may guide us toward these communities.
To assess whether our formulation of community detection matches our needs, we perform extrinsic evaluation against a known ground truth.
While our focus is community detection, our results are sufficiently general for any mesoscopic pattern discovery on graphs.

\section{Preliminaries}

\subsection{Partition structures}
A \emph{partition} or \emph{clustering} $\calC$ of set $X$ can be written as $\calC = \{ C_1, \cdots, C_k \}$, where each \emph{community} or \emph{cluster} $C_i$ is a subset of $X$, their union equals $X$, i.e. $\bigcup_{i=1}^k C_i = X $, and every pair is mutually disjoint, i.e., $\forall$ $i$, $j$, $C_i \cap C_j = \varnothing $.

We reuse some terminology on partition structures from \citet{kingman1978representation}. A \emph{partition} $\Lambda$ of integer $N$ is a multiset of integers $\{ \lambda_1, \cdots, \lambda_k \}$, whose sum is $N$, commonly arranging the elements in descending order. A partition $\calC = \{C_1, \cdots, C_k\}$ of set $X$ is said to have $ \mathrm{shape}(\calC) \triangleq \{ |C_1|, \cdots, |C_k| \}$ (a multiset), where $\sum_{i=1}^k |C_i| = |X| = N$.\footnote{This shape has alternatively been called a \emph{class} or \emph{decomposition pattern} \citep{hauer2016decoding}.} That is, the shape is an integer partition whose elements map to cluster sizes.

\subsection{Random models}

And what, in fact, is the domain of \(\calC\), which we call \(\mathbb{M}\)? For community detection, it will be the set of all partitions on \(N\) nodes. After all, any of these partitions represents a valid community structure, even if it is a poor community structure by intrinsic metrics like modularity. Nevertheless, other random models are relevant to our discussion, restricting the set to those with a fixed number of communities or with a fixed sequence of sizes.

Previous work argued that clustering  similarity should be computed in the context of a random ensemble of clusterings \cite{vinh2009information,hubert1985comparing}. What context of clusterings should be chosen? \citet{gates2017impact} argue that the question above is usually ignored in CD research and more broadly. To remedy this, they proposed three \emph{random models}: $\model{perm}$, $\model{all}$ and $\model{num}$. Given a clustering $\calC$ over set $X$ whose size is $N$, these random models are:

\begin{enumerate}

    \item $\model{all}$: This is the random model that spans all clusterings of $N$ elements: $\model{all}(\calC) = \{ \calC^\prime \mid \sum_{C \in \calC^\prime} |C| = N \}$. This is the random model that we advocate instead of the more common $\model{perm}$; our justifications will be elaborated below.
    
    \item $\model{perm}$ (\emph{permutation model}): The \emph{shape} of the clusterings are fixed, and all random clusterings are generated by shuffling the elements between the fixed-size clusters. Formally, $\model{perm}(\calC) = \{ \calC^\prime  \mid {\rm shape}(\calC^\prime) = {\rm shape}(\calC) \}$. However, despite being widely used for evaluation \citep{zhang2015evaluating,zhang2015relationship,lai2016corrected,vinh2009information,vinh2010information}, the premises of the permutation model are frequently violated; in many clustering scenarios, either the number of clusters or the size distribution vary drastically \citep{hubert1985comparing,gates2017impact}.
    
    \item $\model{num}$: The random model that contains all clusterings of the same number of clusters: $\model{num}(\calC) = \{ \calC^\prime \in \model{all}(\calC) \mid |\calC^\prime| = |\calC| \} $.
\end{enumerate}

These sets of clusterings satisfy the following containment order: $\model{perm}(\calC) \subseteq \model{num}(\calC) \subseteq \model{all}(\calC)$.\footnote{This abuse of notation blurs the distinction between random models and the spaces over which they define their probabilities. We have restricted ourselves to uniform distributions over the spaces, so we comfortably believe that the abuse will not confuse.}

\subsection{Information theory}

After a method has followed clues through the space of partitions, we want to pull off its blindfold to see how it did. Classification accuracy or $F_1$ score won't cut it: When comparing to the ground truth, there are no specific labels (e.g.\ no notion of \say{Cluster 2})---only groups of like entities. We settle for a measure of similarity in the groupings, quantifying how much the computed partition tells us about the ground truth. A popular choice is the \emph{normalized mutual information} (NMI) between the prediction and the ground truth. To understand it (and its flaws), we must review basic information theory concepts.

Mutual information depends on an understanding of \emph{entropy}, which captures the uncertainty in a random variable---in this case, the category labels. For a clustering \(\pred = \{C_1, C_2, \dots, C_k\} \), the entropy is
\begin{equation}
H(\pred) = - \sum_{C \in \pred} \Pr(C) \log \Pr(C)
\text{,}
\end{equation}
%where the entropy of a given cluster is the standard entropy formula:
%\begin{equation}
%H(C) = \sum_{c \in C} \Pr(c \in C) \log \Pr(c \in C)
%\text{.}
%\end{equation}

For community detection, we compute it with maximum likelihood estimation. The probability of membership in a given community is proportional to its size, so the clustering's entropy---a measure of its uncertainty---is 
\begin{equation}
H(\pred) = -\sum_{C \in \pred} \frac{\size{C}}{N} \log \frac{\size{C}}{N} \ .
\end{equation}

Mutual information (MI) measures how well knowing one distribution shrinks our uncertainty about another. Again using maximum likelihood estimation, the MI between a clustering \pred{} and its \say{ground truth} clustering \true{.} is 
\begin{equation}
\label{eq:mutual-information}
I(\pred, \true) = \sum_{C \in \pred} \sum_{T \in \true} \frac{\size{C \cap T}}{N} \log \frac{N \size{C \cap T}}{\size{C}\size{T}}
\text{.}
\end{equation}
It can be understood as the Kullback--Leibler divergence from the joint distribution \(\Pr(C_c, T_t)\) to the product of its marginal distributions \(\Pr(C_c) \Pr(T_t)\). Like entropy, the value is nonnegative. It can be normalized by dividing by an upper bound. We will contrast the choices for this bound in \autoref{sec:recipe}.

\section{What We've Been Doing Wrong} \label{sec:flaws}

Community detection is historically evaluated with NMI:
% \begin{equation}
% \label{eqn:nmi}
% \nmi_{\mathit{\mathcolor{magenta}{sqrt}}}(\pred, \true) = 
% \frac%
% 	{\mi(\pred, \true)}%
% 	{\mathcolor{magenta}{\sqrt{\mathcolor{black}{H(\pred)} \cdot \mathcolor{black}{H(\true)}}}}
% \textrm{.}
% \end{equation}
\begin{equation}
    \label{eqn:nmi}
    \nmi(\pred, \true) = \frac{\mi(\pred, \true)}{\bound(\pred, \true)} \ .
\end{equation}
The quantity is unitless: by normalizing, we divide nats by nats or bits by bits, based on our choice of logarithm. \citet{vinh2009information} and \citet{zhang2015evaluating} both note a \emph{finite size effect}: the average score creeps upward with the number of \emph{predicted} clusters, regardless of the true number. This biases results toward the prediction of a large number of clusters, which is a danger to adequate CD evaluation. A related flaw noted by \citet{peel2017ground} is that the measure is not \emph{homogeneous}. Much the center of a circle will be closest on average to each point in it, the trivial partition into singleton communities scores highest under NMI, when averaged over all possible ground truths. As a homogeneous measure is a precondition of the No Free Lunch theorem \citep{wolpert1996lack}, NMI in fact awards \say{free lunch} when guessing the singletons partition.
We will later show practical consequences of this deficiency.

\section{Recipe for Proper Evaluation} \label{sec:recipe}

Our recommended solution incorporates both practical scoring concerns and an improvement in probability notions over the de facto preference. Desires for an adjustment for chance to give a constant baseline and a constant top score are longstanding, as shown by the popular adjusted Rand index \citep{hubert1985comparing}. The final recommendation is the adjusted mutual information (AMI), computing its expectation with the one-sided random model \modelone{all}. % We denote the expectation operator with random model $\model{perm}$ or $\model{all}$ over a clustering $\pred$ as $\Eperm_\pred$ or $\Eall_\pred$ -- i.e. we define $\Eperm_\pred[f(\pred)] \triangleq \E_{\pred^\prime \sim \model{perm}(\calC)}[f(\pred^\prime)]$. This notation also generalizes to other random models. 

To be useful to practitioners, an evaluation measure for community detection should have the properties of a constant baseline and a constant top score.

\paragraph{Constant baseline} In the statistical sense, we would like a consistent baseline: A random guess should merit no credit. This differs from the notion of a baseline in NMI, which is a lower bound. With this consistent baseline, we eliminate the \say{free lunch}. The proposal of \citet{zhang2015evaluating}, \emph{relative normalized mutual information} ($\rnmi$), improves NMI by subtracting the expected NMI for this ground truth, such that a random guess garners a score of 0. In $\rnmi$ the expected NMI is computed under the permutation model ($\pred^\prime \sim \model{perm}(\pred)$):
\begin{equation}
\rnmi(\pred, \true) = \nmi(\pred, \true) - \E_{\pred^\prime}\left[\nmi(\pred^\prime, \true)\right] \ .
\end{equation}
 % Baseline in the sense of no effort, not in the sense of a lower bound. This measure works with the NFL theorem.

\paragraph{Constant top score} The flaw of rNMI is that now only one reference point (the expectation) is fixed across clusters, not two like in NMI\@. This means that we can't compare performance across clusterings of different sizes, and we don't know whether we've succeeded by attaining the maximum value. \citet{zhang2015relationship} renormalize rNMI to create \emph{renormalized relative normalized mutual information} ($\rrnmi$):
\begin{equation}
\rrnmi(\pred, \true) = \frac{\rnmi(\pred, \true)}{\rnmi(\true, \true)} \ .
\end{equation}
Now, we have both a constant expectation and a constant ceiling, remedying the flaw we've noted.

\paragraph{Symmetry} While we take the controversial stance that symmetry in the measure is undesirable, it is necessary to contextualize additional work, and for showing the connections between methods in \autoref{sec:relationships}. Note that rrNMI's denominator depends only on \(\true\). The information, though, of one distribution about another is not asymmetric. \citet{lai2016corrected} symmetrize rrNMI to create the \emph{corrected normalized mutual information} ($\cnmi$):
\begin{equation}
\cnmi(\pred, \true) = \frac{\rnmi(\pred, \true) + \rnmi(\true, \pred)}{\rnmi(\pred, \pred) + \rnmi(\true, \true)} \ .
\end{equation}
Another measure, \emph{adjusted mutual information} (AMI) \citep{vinh2009information,vinh2010information}, was devised years earlier and incorporates all of these fixes in the style of the adjusted Rand index \citep{hubert1985comparing}: \footnote{The bound function in this AMI definition is $\bound(\pred, \true) = \max_{\pred^\prime, \true^\prime}I(\pred^\prime, \true^\prime)$. In practice we could use any of the upper bounds as described in \citep{gates2017impact}, for example \autoref{eqn:bound}, as long as it is a upper bound consistent with the chosen random model (here $\model{perm}$).}
\begin{equation}
\ami(\pred, \true) = \frac{I(\pred, \true) - \E_{\pred^\prime, \true^\prime}\left[I(\pred^\prime, \true^\prime)\right]} {\displaystyle\max_{\pred^\prime, \true^\prime} I(\pred^\prime, \true^\prime)  - \E_{\pred^\prime, \true^\prime}\left[I(\pred^\prime, \true^\prime)\right]} \ ,
\end{equation}
\noindent where the variables $\pred^\prime \sim \model{perm}(\pred)$ and $\true^\prime \sim \model{perm}(\true)$ under the $\max$ and $\E$ operators above.

\subsection{Probability improvements} \label{sec:probability}
\paragraph{Adjustment considers all relevant partitions}
All of NMI's successors share a fault. Their expectations are computed by incorrectly considering only partitions with the same \emph{decomposition pattern}, or bag of cluster sizes. These aren't the only partitions that could exist, though; the space of options is larger. A community detection algorithm could partition two nodes into either \(\{1\}, \{2\}\) or \(\{1, 2\}\), and ignoring that fact makes assumptions about the structure of the possibility space. 
%\citet{gates2017impact} named some alternate random models for clustering. They called the historical one \model{perm} because it only considers \emph{permutations} of the nodes under a fixed decomposition pattern. For us, the relevant one is \model{all}: all clusterings on \(N\) nodes. 
We should prefer \model{all}.
Fortunately, simple closed-form approximations of its expectation are known \citep{gates2017impact}.

% The number of partitions according to \model{perm} is given by the multinomial theorem: \(\frac{\binom{N}{k1, k2, \dots, k_n}}{k_1 ! k_2 ! \dotsm,  (N - k_1)! (N - k_2) ! \dotsm}\), divided by the number of ways to arrange the groups.

\paragraph{Adjustment recognizes that the ground truth is fixed.}
Fixing the first fault in our symmetric measures leaves a second fault. We compute the expectation over all clusterings \emph{and} all truths in the baseline $\baseline(\pred, \true)$. But the truth for our problem is fixed. A different truth means we're solving a different problem---core--periphery, etc. Our expectations should only ever be over the values of \(\pred\):
\begin{equation}
    \label{eqn:good-baseline}
    \baseline_{\rm all}^1(\pred, \true) = \E_{\pred^\prime \sim \model{all}(\true)}[\mi(\pred^\prime, \true)] \ .
\end{equation}
This all too is easy to plug into our formulas, as straightforward closed-form approximations again exist \citep{gates2017impact}.

The textbook AMI (before these probability improvements) is known to be asymptotically homogeneous, imparting negligible advantage for weak guesses in the limit as \(N\) approaches infinity \citep{peel2017ground}. By correcting for chance properly, we can make this asymptotically diminishing advantage exactly zero \citep{mccarthy2018exact}.

\begin{figure}
	\centering
	\includegraphics[width=\linewidth]{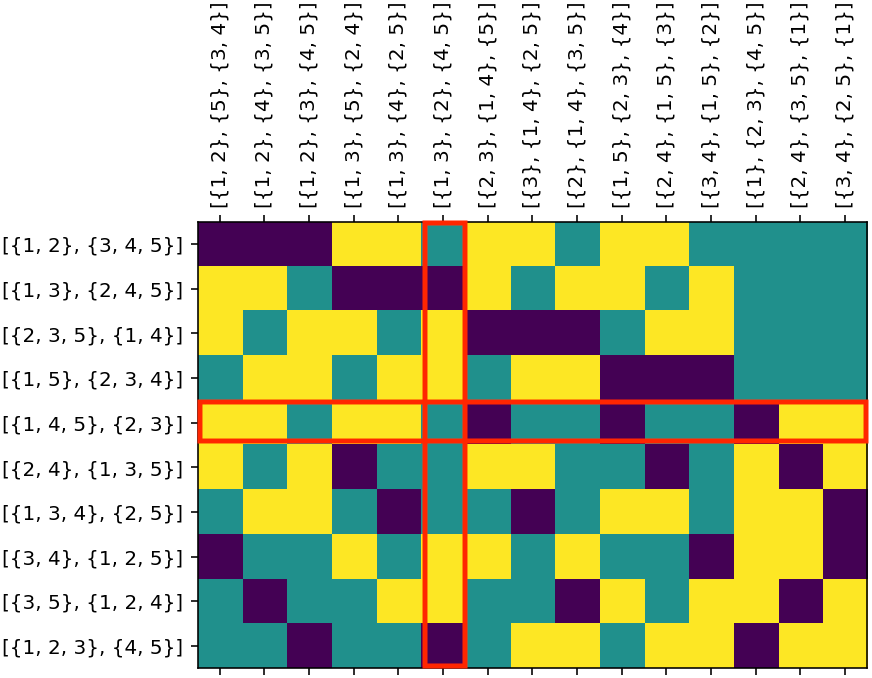}
	% \missingfigure[figwidth=\linewidth]{Expectation matrix}
	\caption{Heatmap of mutual information between two partitions. AMI computes expectation over all \((X', Y')\) pairs in the joint distribution. cNMI examines only the observation's row and column. In \(\mathbb{M}_\mathrm{perm}\) (shown), exchangeability under MI makes these expectations equal.}
%MI matrix for the permutation randomness model. In the MI matrix, entries are indexed by pairs of partitions drawn from the same distribution as the observation. AMI computes expectation over the \emph{entire} matrix. cNMI examines only the observation's row and column. Label exchangeability makes these values equal.
	\label{fig:expectation_matrix}
\end{figure}

\subsection{Bonus: Generalized mean} \label{sec:mean}

There's a final component to our recommendation: How do we compute the upper bound function $\bound(\pred, \true)$? Traditionally, different generalized means of the cluster entropies have been used (e.g., \autoref{eqn:bound} uses the geometric mean) \citep{vinh2010information}. \citet{yang2016comparative} have discussed that the particular choice of generalized mean is unimportant. Fortunately, the problem is simpler in our case: \citet{gates2017impact} discuss that in \model{all}, the bounding entropy for either cluster is \(\log N\), and any generalized mean of \(\log N\) and \(\log N\) is \(\log N\).\footnote{By the generalized mean inequality.} Hence we define our upper bound function in $\model{all}$ to be
\begin{equation}
    \label{eqn:good-bound}
    \bound_{\rm all}(\pred, \true) = \log N \ .
\end{equation}

\section{Relationships Between Measures} \label{sec:relationships}

Having waded through the alphabet soup of information-theoretic measures, we now present our major theoretical result: explorations of interrelations between measures. While we began this with our discussion of generalized means, these relationships exploit the random models and sidedness from \autoref{sec:recipe}. 

\begin{figure}[t]
  \centering
  \includegraphics[width=0.8\linewidth]{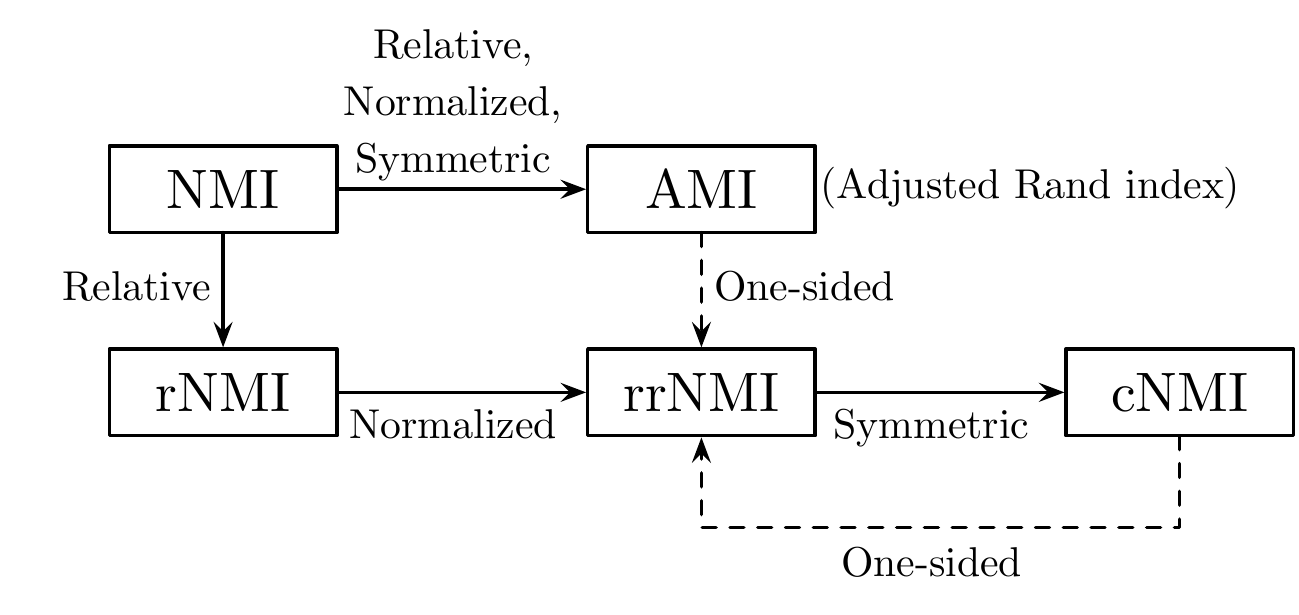}
  \caption{Relationship between various information-based CD measures.}
  \label{fig:relationship}
\end{figure}

The relationships are summarized in \autoref{fig:relationship}. To characterize these relationships, we rely on the notion of function \emph{specialization}: limiting a function's expressiveness by fixing some parameters, altering its behavior \citep{young2018universality,veldt2018unifying}.  As the simplest example, one-sided measures in general specialize their two-sided counterparts by fixing the ground truth \true, rather than taking an expectation over \true's universe.

Our first finding is that in \model{perm}, AMI specializes itself: One-sided and two-sided AMI are equivalent. We demonstrate this using the exchangeability of partitions under permutation in \autoref{sec:rachel}. For a graphical intuition, we redirect the viewer to \autoref{fig:expectation_matrix}.

Building on this, we also show that rrNMI (an inherently one-sided, asymmetric measure) specializes AMI in \model{perm}. We exploit a clever form of \(1\): Dividing the numerator and denominator of rrNMI both by the maximum MI renders the expression identical to one-sided AMI. This does not generalize to other random models because it relies on the exchangeability of partitions.

Oddly, rrNMI \emph{also} specializes cNMI into one-sided cNMI in all random models. This is a straightforward algebraic result. It becomes clear by expressing both measures in terms only of mutual information and its expectation. 

This may lead the reader to suspect that cNMI and AMI are equivalent, at least under some random model. Unfortunately, under all three of the \citet{gates2017impact} models, there are irremediable differences. In \model{perm}, the denominators compute expectations over different sets. In \model{all} and \model{num}, the lack of exchangeability renders the numerators distinct.

As a final conceptual introduction, we review the mediant, or \say{freshman sum}, of two fractions. It is given by separately adding the numerators and denominators of two fractions. It has the property that it lies between its two arguments. The cNMI symmetrizes rrNMI as the mediant of each one-sided measure, so we know that the cNMI is always bounded between the one-sided values. Finally, because the numerators are identical in each one-sided variant, the mediant is also the harmonic mean; this can be shown through simple algebraic manipulation.

Our recommended measure, \textbf{one-sided AMI in \model{all}}, is identical to rrNMI in \model{all}. Though rrNMI is already closer to an appropriate community detection evaluation, we follow parsimony and historical precedent to use the name AMI. 

Therefore our recommended measure, the one-sided AMI in $\model{all}$, by combining our derived baseline function $\baseline_{\rm all}^1$ in \autoref{eqn:good-baseline} and upper bound function $\bound_{\rm all}$ in \autoref{eqn:good-bound}, could be formally written as
\begin{equation}
    \label{eqn:final-ami}
    \ami_{\rm all}^1 (\pred, \true) = \frac{\mi(\pred, \true) - \E_{\pred^\prime}{[\mi(\pred^\prime, \true)]}} { \log N - \E_{\pred^\prime}{[\mi(\pred^\prime, \true)] }} \ ,
\end{equation}
where $\pred^\prime \sim \model{all}(\true) = \model{all}(\pred)$.
\section{A View to a Trap}

Having given the design of a fitting extrinsic evaluation, we now extol its need.
Community detection methods can be viewed, at their core,  simply as efficient heuristics for maximizing modularity while remaining tractable.
In choosing a method, we first select graphs which we believe represent the distribution of graphs in our use-case.
We then apply our methods to the graphs, scoring their predictions against the ground truth and choosing the method with best performance.
(We may also factor running time into our decision---a cost-aware objective.)
But NMI's scores are misleading, making it a danger to our method selection process.

To showcase the danger, we set and spring a trap for NMI\@.
We use a community detection method with a unique architecture, making it susceptible to choosing trivial clusterings.
This elicits NMI's pathology: 
NMI exaggerates the method's performance compared to other methods, awarding credit to the trivial clustering.
AMI corrects this exaggeration.

\section{The Trap: Leximin, An Adversarial Method}
\begin{figure}
	\centering
    \includegraphics[width=6cm]{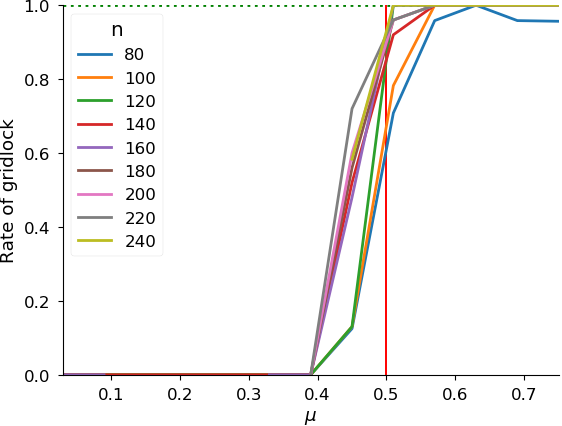}
    \caption{The rate of gridlock for various mixing parameters~$\mu$ and network sizes. 25 LFR graphs were sampled for each combination of parameters.}
    \label{fig:gridlock}
\end{figure}

The \emph{leximin method} is a divisive clustering method, motivated by congestion in traffic networks. It formulates community detection as a hierarchical version of the \emph{sparsest cut problem}, an objective related to modularity. (Notably, both reduce to \textsf{MAX-CUT}\@ \citep{matula1990sparsest}, making them \textsf{NP}-hard. \citet{veldt2018unifying} show that both objectives specialize an underlying function.)

The sequence of cuts comes from a linear programming (LP) relaxation of the sparsest cut problem: the \emph{maximum concurrent flow problem} \citep{shahrokhi1990maximum}. 
The model routes traffic flow between all pairs of nodes, with an objective that minimizes \emph{congestion} while fairly satisfying demand \cite{shahrokhi1990maximum}.
(In applications, the traffic could be goods in a commodity distribution network, gossip in a social network, or signals in a neural network.)
When rerouting cannot avoid congestion, the saturated edges define a sparse \say{bottleneck cut}.
Continuing to increase the allocation of flow gives a sequence of bottleneck cuts which dissect the graph \cite{matula1990sparsest}.

Or at least, that would be nice.
Lamentably, the tractability comes from \emph{weak} duality between hierarchical sparsest cut and the hierarchical \emph{maximum concurrent flow problem}.
The relaxation will either find the sparsest cut or a multipart cut (a \emph{grid}) when the maximum throughput is less than the density of any cut.
This is consistent with LPs being in the complexity class \textsf{P} while sparsest cut is \textsf{NP}-hard.
The competing forces of cut density and the gridlock bound mean that the graph may splinter into single-node communities without any mesoscopic structure in between.
We exploit the fact that complete gridlock becomes reliable for graphs of certain structural properties examined below, as shown in \autoref{fig:gridlock}.

\section{Springing the Trap: Experiment}

\def\subfigscale{0.3}
\begin{figure*}
\centering
\subfloat[NMI, Multilevel (Louvian)]{%
  \includegraphics[width=\subfigscale\linewidth]{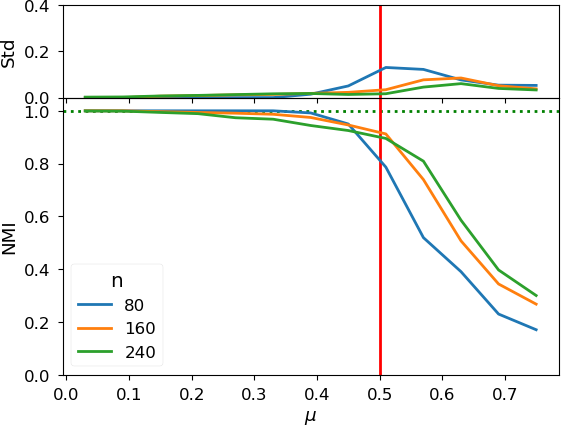}%
}~
\subfloat[NMI, Edge betweenness]{%
  \includegraphics[width=\subfigscale\linewidth]{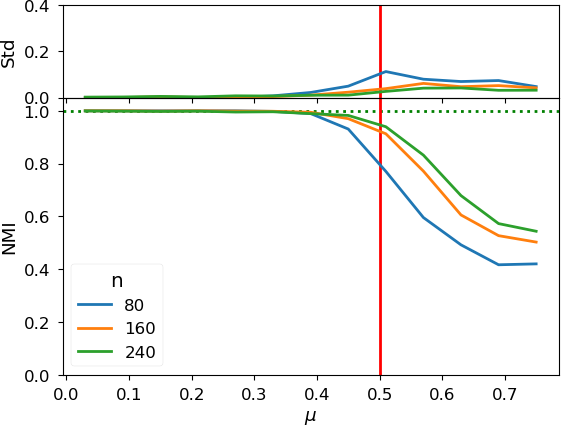}%
}~
\subfloat[NMI, \textbf{Leximin}]{%
  \includegraphics[width=\subfigscale\linewidth]{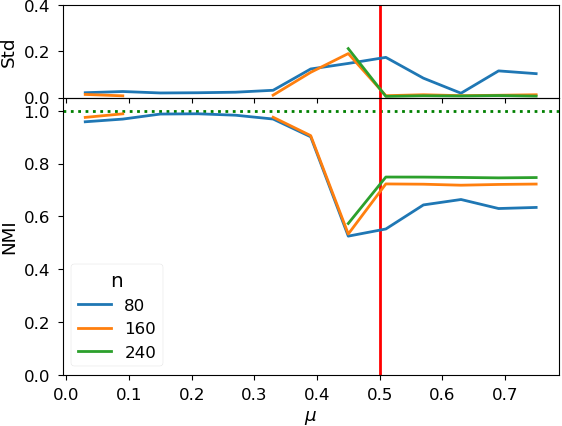}%
}\\%

\subfloat[AMI, Multilevel (Louvian)]{%
  \includegraphics[width=\subfigscale\linewidth]{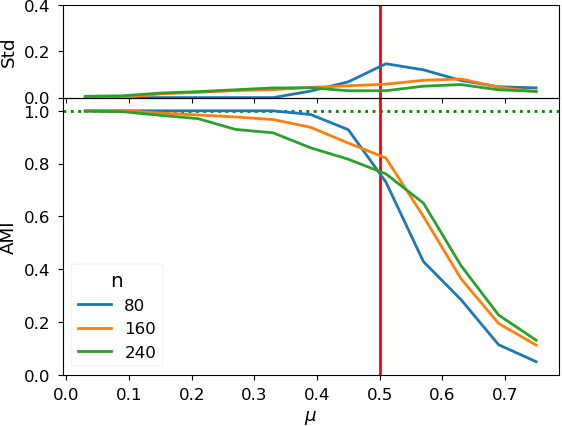}%
}~
\subfloat[AMI, Edge betweenness]{%
  \includegraphics[width=\subfigscale\linewidth]{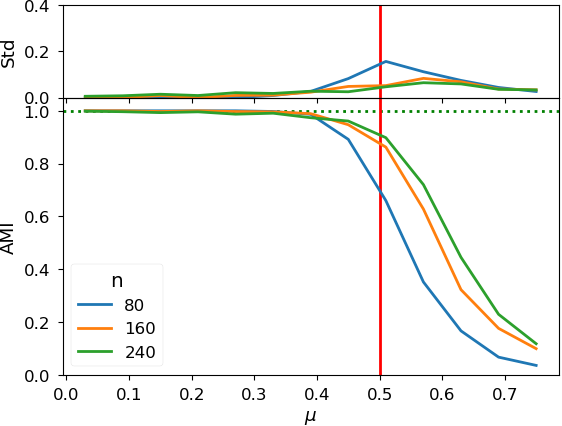}%
}~
\subfloat[AMI, \textbf{Leximin}]{%
  \includegraphics[width=\subfigscale\linewidth]{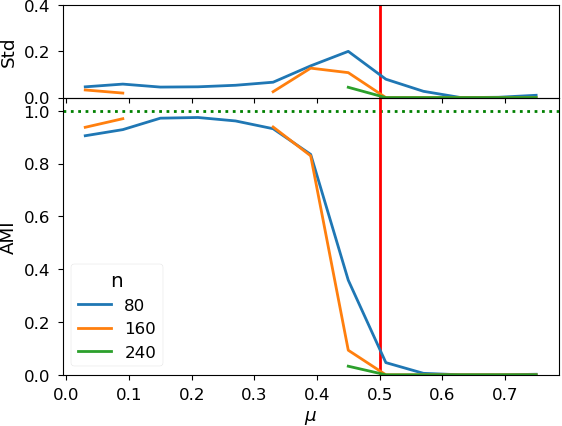}%
}\\%

\caption{The mean and st.\ dev.\ of NMI (top) and AMI (bottom) as a function of the mixing parameter, \(\mu\).
The colors represent different network sizes.
The vertical line at \(\mu = 0.5\) divides strong and weak communities \citep{radicchi2004defining}, and the dotted line shows the maximum score, 1.0.}
\label{fig:results}       % Give a unique label
\end{figure*}%

With the trap in place, we now spring it with an experiment of the fashion described above. We benchmark the NMI and AMI of eight popular CD methods, plus the adversarial case: the leximin method. As our graph distribution, we chose the Lancichinetti--Fortunato--Radicchi (LFR) benchmark graphs, which manifest a ground truth and obey properties of real-world networks \citep{lancichinetti2008benchmark}.

Following \citet{yang2016comparative}, we test \(25\) LFR realizations for each combination of parameters \(N \in \{80, 100, \dots, 240\}\) and \(\mu \in \{0.03, 0.09, \dots, 0.75\}\). Here, \(\mu\) is the \emph{mixing parameter}. It controls the fraction of each node's edges that connect outside of its community, and it can be thought of as a knob to increase the difficulty of the community detection task. All other parameters were set as in \citet{yang2016comparative}. We report the mean and standard deviation of each score for each combination of parameters.

The eight methods we compare against are: Fastgreedy~\cite{clauset2004finding}, Infomap~\cite{rosvall2007information, rosvall2009map}, label propagation~\cite{raghavan2007near}, leading eigenvector~\cite{newman2006finding}, multilevel~\cite{blondel2008fast}, spinglass~\cite{reichardt2006statistical}, and walktrap~\cite{pons2005computing}. For a concise description of each method, the reader is directed to \citet{yang2016comparative}.

\paragraph*{Implementation details} The leximin method is implemented in \texttt{AMPL}~\cite{fourer1989mathematical}. While the hierarchical MCFP could be expressed as a single LP\@, we exploit the lexicographic problem structure to decompose the problem into a sequence of \(N - 1\) smaller, subordinate LPs \citep{podinovskii1972lexicographical}. Using \texttt{AMPL}'s \texttt{CPLEX} backend, the method solves the LP of \citeauthor{dong2015compact}'s \emph{triples} formulation of the MCFP~\cite{dong2015compact}. The eight other algorithms are implemented in the \texttt{igraph} package, accessed through \texttt{igraph}'s R interface. Determining optimal modularity is built-in for the \texttt{igraph} algorithms; it is done using the \texttt{networkx} Python library~\cite{hagberg2008exploring} for leximin. Cluster evaluation is done for all using the NMI and AMI functions of the \texttt{scikit-learn} Python package~\cite{pedregosa2011scikit}.

\section{Postmortem of a Trapped Measure: Discussion and Results}

The most useful graph characteristic to capture the performance of a community detection algorithm is the strength of its communities \citep{yang2016comparative}. One proposed distinction of communities is into \say{strong} and \say{weak} \citep{radicchi2004defining}. Strong communities are tight-knit: their nodes all have more edges to each other (intra-community) than to other communities' nodes (inter-community). In the LFR generator, the strength of communities is controlled \emph{en masse} by the mixing coefficient, \(\mu\). Values closer to \(1\) indicate a harder problem, edging us toward a \emph{detectability limit} beyond which no method (real or hypothetical) can identify communities.

We present the mean and standard deviation of our scores in \autoref{fig:results}. Those graphs which took longer than \(3\) hours to process are excluded, though all graphs with \(n \geq 180\) and \(\mu \geq 0.5\) were processed within this bound. The surprising fact that larger graphs were processed faster comes from our column generation scheme: When gridlock splinters the graph into singletons at an early stage, we've reached our optimum, and the remaining \(O(N)\) LPs need not be solved---a tremendous reduction in computational labor.

Cranking the mixing coefficient \(\mu\) higher makes communities weaker. Our tests revealed, as expected, that the leximin method's NMI begins to surpass the others, staying high when they fall. Infomap and Label Propagation plunge down to 0, while Leading Eigenvector and Fastgreedy have gradual downward slopes, never performing nearly as well as the others. More typically, Multilevel, Spinglass, and Walktrap remain high while communities are strong, then fall near the boundary of \(\mu = 0.5\).

These results, though, are very misleading. Remember from \autoref{fig:gridlock} that gridlock occurs almost consistently when we cross the threshold into weak communities. This means that no mesoscopic structure is detected. Why should this be scored higher than methods that extract some of the embedded patterns?

 When partitions are scored using AMI, most methods' performance profiles are unchanged. The Louvian (multilevel) method and Infomap, for instance, continue to perform strongly on strong communities and drop off at the same values of~\(\mu\). The edge betweenness algorithm's performance drops faster as \(\mu \rightarrow 1\), but its trend is similar to the trend under NMI. The same is true for Fastgreedy method. 
By contrast, AMI gives a completely different characterization of the leximin method's performance. Against the backdrop of this relative consistency, we see a dramatic drop in the performance of the leximin method, lining up with the increased rate of predicting singleton clusterings. Leximin's score for high \(\mu\) is zero. This is a fairer assessment of the methods' ability, matching our intuition about how good one clustering should be compared to another---particularly, the fact that the singleton clustering finds no community structure, so we should regard this as a failure when community structure does exist.

After a broad study on factors affecting community detection, \citet{yang2016comparative} proposed a method for choosing a community detection algorithm. Based on runtime and NMI\@, the authors recommended certain community detection algorithms for particular combinations of \(N\) and \(\mu\). If we followed their method precisely, then the leximin method would be recommended for small graphs with weak communities. With the truth quantified by AMI\@, we see that leximin should only be used for networks with strong communities, and its runtime limits its application to small to medium-sized graphs.

\section{Related Work}
Over the years, numerous evaluation methods for community detection have been suggested. While NMI has historically been the most popular (first proposed for community detection by \citet{danon2005comparing}), others include the variation of information \citep{meila2003comparing}, the V-measure \citep{rosenberg2007vmeasure} (a rediscovery of NMI), the adjusted Rand index (ARI) \citep{hubert1985comparing}, F-score \citep{dhillon2003divisive}, and Cohen's \(\kappa\) \citep{liu2018evaluation}.

The standardized mutual information \citep{romano2014standardized} was proposed as another adjustment for chance. Using instead the variance as the denominator, it no longer has the constant top score we seek. \citet{romano2016adjusting} identified both AMI and ARI as specializations of a general function. Finally, \citet{decelle2011asymptotic} proposed a measure called overlap that is adjusted for chance. We favor AMI because it is most similar to the conventional measure and also meets our desiderata.

\section{Conclusion and Future Work}

We see that the leximin method, like many, is successful on strong communities and is incorrectly appraised by NMI. 
While NMI was the best known measure when \citeauthor{danon2005comparing} first used it for CD~\cite{danon2005comparing}, we see that \textbf{NMI can exaggerate community detection performance}.
Inertia in the CD community is likely why a departure from NMI has not occurred.
Nevertheless, the measure's flaws demand a move to a more robust measure.
The choice of AMI corrects this exaggeration, so we encourage AMI's use in the CD community moving forward.

The curious reader may note that mutual information is already a symmetric measure. An asymmetric measure like relative entropy makes a promising avenue for future work. Another important avenue is finding fair and appropriate measures for hierarchical or overlapping clustering~\citep{horta2015comparing}. Finally, future work will assess CD methods and the related \(k\)-way partition problem using one-sided AMI under \(\mathbb{M}_\mathrm{all}\) and \(\mathbb{M}_\mathrm{num}\) respectively, as well as varying the generalized mean parameter \(p\).

\bibliographystyle{apsrev4-1}
\bibliography{wns-bib}

\appendix

\section{Self-specialization of AMI}
\label{sec:rachel}

We show that the one-sided AMI and two-sided AMI are identical under \model{perm}---that AMI specializes itself. Superficially, the differences between the two equations are in the expectation and the max-term---both are over different distributions. By showing that each of these are in fact identical between the equations, we will show that AMI self-specializes. For this, we will discuss clusterings $\calC$ and $\calT$ on set $X$.

The upper bound function is straightforward; it is some generalized mean \(M_p\) of the two clusterings' entropies \(H(\calC)\) and \(H(\calT)\). This bound is unchanged whether using one-sided or two-sided AMI, because the entropy of a clustering depends only on its decomposition pattern---not the actual elements. So much for the upper bound function.

Now, we must show that the expectation is the same regardless of whether one or both variables is bound.\footnote{The reader may recall \autoref{fig:expectation_matrix} if a concrete version of the problem is helpful.} Formally,
\begin{align*}
  \E_{\calC^\prime, \calT^\prime} [\mi(\calC^\prime, \calT^\prime)] &= \E_{\calC^\prime} [\mi(\calC^\prime, \calT)] \\
\end{align*}
\noindent given $\calC^\prime \in \model{perm}(\calC)$ and $\calT^\prime \in \model{perm}(\calT)$. We note that either variable may be fixed due to the symmetry of mutual information, and the claim will still hold.

We will first show that the expectation is impervious to permutations of the input, then we will use this resilience to show that the expectation is the same for one- or two-sided expectations.

%An alternative formula reveals that it is the sum of the two partitions' entropy, minus their \emph{joint entropy}: \(I(\pred, \true) = H(\pred) + H(\true) - H(\pred, \true)\).

% \begin{proof}
  
%   We first prove the following lemma on permuting the elements in the set.
%   \begin{lemma}
%     For any permutation $\pi$ on set $X$, we have 
%     \begin{equation*}
%       \E_{\calC^\prime \sim \model{perm}(\calC)} [F(\calC^\prime, \calS)] = \E_{\calC^\prime \sim \model{perm}(\calC)} [F(\pi(\calC^\prime), \calS)] \ .
%     \end{equation*}
%   \end{lemma}
  
%   \begin{proof}[Proof of the lemma]
%     \begin{align*}
%       \pi(\model{perm}(\calC)) & \triangleq \{ \pi(\calC^\prime) \mid \calC^\prime \in \model{perm}(\calC) \} \\
%       &= \{ \calC^\prime \mid \calC^\prime \in \model{perm}(\calC) \} \\
%       &= \model{perm}(\calC)  \ .
%     \end{align*}
%     Hence, 
%     \begin{align*}
%       \E_{\calC^\prime \sim \model{perm}(\calC)} [ F(\calC^\prime, \calS) ] 
%                                       &= \E_{\calC^\prime \sim \pi(\model{perm}(\calC))} [F(\calC^\prime, \calS)] \\
%                                       &= \E_{\calC^\prime \sim \model{perm}(\calC)} [F(\pi(\calC^\prime), \calS)] \ . \qedhere
%     \end{align*}
%   \end{proof}  

It is clear from the definition of mutual information (\autoref{eq:mutual-information}) that it is permutation-invariant: it relies on intersections and sizes of sets, and in aggregate these are unaltered by permuting the labels.

Now, we rely on the following theorem:

\begin{theorem}[\citet{schumacher2001no}]
  For any set $X$, given partition $\calT_1$ and $\calT_2$ on $X$ with the same shape,  another partition $\calC$ on $X$ (not necessarily the same shape as $\calT_1$), and any permutation-invariant function $F(\cdot, \cdot)$ on pairs of partitions, we have
  \begin{equation*}
    \E_{\calC^\prime \sim \model{perm}(\calC)}[F(\calC^\prime, \calT_1)] = \E_{\calC^\prime \sim \model{perm}(\calC)}[F(\calC^\prime, \calT_2)] \ .
  \end{equation*}
\end{theorem}

This amounts to saying that the expectation is independent of the particular fixed variable (\(\calC\) or \(\calT\)) so long as the fixed variable is drawn from \model{perm}. That is:
\begin{equation}
    \E_{\calC'} [\mi(\calC', \calT_1)] = \E_{\calC'} [\mi(\calC', \calT_2)] = K
    \text{.}
\end{equation}

This value is a row average in \autoref{fig:expectation_matrix}; more broadly, it is a one-sided expectation for the randomness model \model{perm}. What remains is to show that this average matches the two-sided average, which relies on simple algebra.
\begin{align*}
    \E_{\calC', \calT'} \left[I(\calC', \calT')\right] &= \E_{\calT'} \left[ \E_{\calC'} \left[I(\calC', \calT')\right] \right] \\
    &= \E_{\calT'} [K] \\
    &= K \ .
\end{align*}

Thus, the two-sided expectation matches the one-sided expectation. \hfill \qedsymbol

\end{document}